\documentclass[a4paper]{article}
\RequirePackage[english]{babel}
\RequirePackage[latin1]{inputenc}
\RequirePackage[T1]{fontenc}
\RequirePackage{mathrsfs}
\RequirePackage{amsmath}
\RequirePackage{amssymb}
\RequirePackage{amsbsy}
\RequirePackage{bm}
\begin{document}
\title{\bf{Geometrical Properties and Propagation\\ for the Proca Field Theory}}
\author{Luca Fabbri\\ 
\footnotesize (INFN \& Theory Group, Department of Physics, Bologna, ITALY)}
\date{}
\maketitle
\ \ \ \ \textbf{PACS}: 04.20.Cv (Fundamental problems and general formalism)
\begin{abstract}
We consider the Proca field with dynamical term given by the exterior derivative with respect to the most general connection; the most general Proca field equations are given, and a discussion about the propagation and the geometrical properties are presented: it is shown that this generalization is inconsistent. So the standard theory is already the most general Proca Theory possible.
\end{abstract}
\section*{Introduction}
As the existence of the massive vector bosons has clearly shown, the Proca vector field is a very important field in physics.

Its importance resides on the fact that, being it a vector field with dynamical term given by the exterior derivative and having mass, it allows for the description of all vector fields that are gauge fields before eventually getting their masses through a mass generation mechanism; another fundamental reason is that, since the dynamical term is the divergence of the curl of the vector field and there is the mass term, then this particular structure of its field equations automatically provides the subsidiary condition that reduces the number of degrees of freedom to those needed to define massive vector fields.

However, the fact that this field has dynamical term written in terms of the antisymmetric part of the derivative, which can be defined without torsion, does not prevent us to try to generalize it up to the the exterior derivative calculated with respect to the most general connection, in which torsion is present in a natural way; this most general connection is commonly not used because as it is known it would spoil the gauge invariance, but there is no gauge invariance to save in this case for this field is massive: therefore such a general connection can be employed.

In this paper we will consider such a theory, deriving its consequences and discussing its implications regarding the Proca field, its propagation and its most important geometrical properties.
\section{Fundamental Definitions}
In a given geometry, the metric structure is given in terms of two symmetric metric tensors $g_{\alpha\beta}$ and $g^{\alpha\beta}$ that are one the inverse of the other, and differential operations $D_{\mu}$ are defined through the connections $\Gamma^{\rho}_{\alpha\beta}$; the metric tensor are to be such that they can be locally reduced to the Minkowskian form of signature $(1,-1,-1,-1)$, and the covariant derivatives applied upon the metric tensors are required to vanish according to what is called metricity condition $D_{\mu}g=0$, as discussed in \cite{h-h-k-n}. Furthermore, requiring this condition of metricity for any connection leads to the complete antisymmetry of Cartan torsion tensor $Q_{\alpha\mu\rho}$, as explained in \cite{f/1}. 

In this background, we will define Riemann curvature tensor $G_{\alpha\beta\mu\nu}$ as
\begin{eqnarray}
G^{\alpha}_{\lambda\mu\nu}=
\partial_{\mu}\Gamma^{\alpha}_{\lambda\nu}-\partial_{\nu}\Gamma^{\alpha}_{\lambda\mu}
+\Gamma^{\alpha}_{\rho\mu}\Gamma^{\rho}_{\lambda\nu}
-\Gamma^{\alpha}_{\rho\nu}\Gamma^{\rho}_{\lambda\mu}
\label{Riemann}
\end{eqnarray}
antisymmetric in both the first and the second couple of indices, allowing only one independent contraction, Ricci curvature tensor $G^{\lambda}_{\alpha\lambda\beta}=G_{\alpha\beta}$, whose contraction is Ricci curvature scalar $G_{\alpha\beta}g^{\alpha\beta}=G$ and this will set our convention.

Riemann curvature tensor, Ricci curvature tensor and scalar, together with Cartan torsion tensor verify 
\begin{eqnarray}
D_{\rho}Q^{\rho\mu \nu}
+\left(G^{\nu\mu}-\frac{1}{2}g^{\nu\mu}G\right)
-\left(G^{\mu\nu}-\frac{1}{2}g^{\mu\nu}G\right)\equiv0
\label{torsiondiv}
\end{eqnarray}
and
\begin{eqnarray}
D_{\mu}\left(G^{\mu\rho}-\frac{1}{2}g^{\mu\rho}G\right)
-\left(G_{\mu\beta}-\frac{1}{2}g_{\mu\beta}G\right)Q^{\beta\mu\rho}
+\frac{1}{2}G^{\mu\kappa\beta\rho}Q_{\beta\mu\kappa}\equiv0
\label{curvaturediv}
\end{eqnarray}
which are geometric identities in the form of conservation laws, called Jacobi-Bianchi identities.

We remark that from the metric tensor it is possible to define the Levi-Civita tensor $\varepsilon$ for which $D_{\mu}\varepsilon=0$ precisely because of the complete antisymmetry of torsion. 

In turn, since torsion is completely antisymmetric then we can write
\begin{eqnarray}
Q^{\beta\mu\rho}=\varepsilon^{\beta\mu\rho\sigma}W_{\sigma}
\label{axialvector}
\end{eqnarray}
in terms of what is called axial torsion vector.

Within this background, to define matter fields that can be classified according to the value of their spin we have to consider that a given matter field of spin $s$ possesses $2s+1$ degrees of freedom, which have to correspond to the $2s+1$ independent solutions of a system of equations that specify the highest-order time derivative for all components of the field, called system of matter field equations.

However, since it may happen that field equations are not enough to determine the correct rank of the solution, restrictions need to be imposed in terms of equations in which all components of the field have highest-order time derivatives that never occur, called constraints; these constraints can be imposed in two ways, either being implied by the field equations, or being assigned as subsidiary conditions that come along with the field equations themselves. 

Although the former procedure seems more elegant, whenever interactions are present it can give rise to two types of problems, the first of which concerning the fact that the presence of the interacting fields could increase the order derivative of the constraining equation up to the same order derivative of the field equations themselves, creating the possibility that highest-order time derivatives of some component occur, converting the constraint into a field equation, then spoiling the counting of degrees of freedom.

Before proceeding we have to remind the reader that to check causal propagation, the general method is to consider in the field equations eventually modified by constraints the terms of the highest-order derivative of the field, formally replacing the derivatives with the vector $n$ in order to obtain the propagator, of which one has to compute the determinant setting it to zero in order to get an equation in terms of $n$ called characteristic equation, whose solutions are the normal to the characteristic surfaces, representing the propagation of the wave fronts: if there is no time-like normal among all the possible solutions, then there is no space-like characteristic surface, and therefore these is no acausal propagation of the wave front. 

If in the constraining equation the highest-order time derivative never appeared, or if it actually appeared but could be removed by means of field equations, then the constraint is a constraint indeed, but in this case a second type of problem can arise, regarding the fact that the interacting fields could let appear terms of the highest-order derivative in the propagator, allowing these terms to influence the propagation of the wave fronts themselves, as it is explained in \cite{v-z}.

Once this analysis is performed, causal propagation of wave fronts is checked, and the exact number of degrees of freedom of the matter field solution is established, the last requirement for this system of matter field equations is that they have to ensure the complete antisymmetry of the spin, so that taking the spin $S^{\nu\sigma\rho}$ with the energy $T^{\sigma\rho}$ they have to be such that the relationships 
\begin{eqnarray}
D_{\rho}S^{\rho\mu\nu}+\frac{1}{2}\left(T^{\mu\nu}-T^{\nu\mu}\right)=0
\label{conservationspin}
\end{eqnarray}
and
\begin{eqnarray}
D_{\mu}T^{\mu\rho}-T_{\mu\beta}Q^{\beta\mu\rho}-S_{\beta\mu\kappa}G^{\mu\kappa \beta \rho}=0
\label{conservationenergy}
\end{eqnarray}
are verified, implying the whole set of field equations
\begin{eqnarray}
\left(G^{\sigma\rho}-\frac{1}{2}g^{\sigma\rho}G\right)=-\frac{1}{2}T^{\sigma\rho}
\label{einstein}
\end{eqnarray}
and
\begin{eqnarray}
Q^{\nu\sigma\rho}=S^{\nu\sigma\rho}
\label{sciama-kibble}
\end{eqnarray}
to be such that the conservation laws (\ref{torsiondiv}) and (\ref{curvaturediv}) are satisfied automatically.

This determines the set-up of the fundamental field equations in minimal coupling, that is taking the least-order derivative possible in both sides of the field equations.
\section{Propagation and Geometrical Properties}
Having settled the background in this way, and because the background is characterized by these restrictions, then matter fields will behave in a correspondingly restricted way, as it is also explained in \cite{f/2}.

Now, we begin to consider the issue of which matter vector fields could possibly be defined within this background.

In the case of a vector $V_{\mu}$ it is possible to define beside the standard covariant derivative given in terms of the connection another most special differential operation given by $Z_{\mu\nu}=\partial_{\mu}V_{\nu}-\partial_{\nu}V_{\mu}$ in terms of no additional field and called curl or exterior derivative, which can be generalized up to the differential operator given by $Z_{\rho\mu}=D_{\rho}V_{\mu}-D_{\mu}V_{\rho}$ that is formally the exterior derivative but now with respect to the most general connection.

So given the vector field $V_{\mu}$, we postulate the most general Proca matter field equations as
\begin{eqnarray}
D_{\mu}Z^{\mu\alpha}
+\frac{\lambda}{2} D_{\mu}Z_{\eta\rho}\varepsilon^{\mu\eta\rho\alpha}+m^{2}V^{\alpha}=0
\label{fieldequations}
\end{eqnarray}
which specify the second-order time derivative for only the spatial components, but which also develop the constraint
\begin{eqnarray}
&m^{2}D_{\mu}V^{\mu}
-\frac{\lambda}{4}Q_{\rho\mu\nu}D^{\rho}Z_{\alpha\beta}\varepsilon^{\alpha\beta\mu\nu}
-\frac{1}{2}Q^{\rho\alpha\beta}D_{\rho}Z_{\alpha\beta}-\\
\nonumber
&-\frac{\lambda}{2}D_{\mu}Q^{\rho}_{\phantom{\rho}\beta\nu}Z_{\rho\alpha}\varepsilon^{\alpha\beta\mu\nu}
-\frac{1}{2}D_{\rho}Q^{\rho\alpha\beta}Z_{\alpha\beta}=0
\label{constraint}
\end{eqnarray}
and where the conserved quantities are given by the energy
\begin{eqnarray}
\nonumber
&T^{\alpha\mu}=
-\frac{1}{2}g^{\alpha\mu}m^{2}V^{2}
+\left(\frac{1}{4}g^{\alpha\mu}Z_{\rho\eta}Z^{\rho\eta}
-Z^{\mu\theta}Z^{\alpha}_{\phantom{\alpha}\theta}\right)+\\
&+D_{\rho}V^{\mu}\left(Z^{\rho\alpha}+\frac{\lambda}{2}Z_{\sigma\theta}
\varepsilon^{\sigma\theta\rho\alpha}\right)
\label{energy}
\end{eqnarray}
and the spin
\begin{eqnarray}
S^{\rho\alpha\beta}=\frac{1}{2}
\left[V^{\alpha}\left(Z^{\rho\beta}+\frac{\lambda}{2}Z_{\sigma\theta}
\varepsilon^{\sigma\theta\rho\beta}\right)
-V^{\beta}\left(Z^{\rho\alpha}+\frac{\lambda}{2}Z_{\sigma\theta}
\varepsilon^{\sigma\theta\rho\alpha}\right)\right]
\label{spin}
\end{eqnarray}
so that, whereas the condition
\begin{eqnarray}
V^{\alpha}\left(Z^{\rho\beta}
+\frac{\lambda}{2}Z_{\sigma\theta}\varepsilon^{\sigma\theta\rho\beta}\right)
+V^{\rho}\left(Z^{\alpha\beta}
+\frac{\lambda}{2}Z_{\sigma\theta}\varepsilon^{\sigma\theta\alpha\beta}\right)=0
\label{condition}
\end{eqnarray}
ensures the complete antisymmetry of the spin, this form of the spin with the energy is such that the conservation laws (\ref{energy}) and (\ref{spin}) are verified. We notice a couple of facts about the constraints (\ref{constraint}) and the condition of complete antisymmetry of the spin (\ref{condition}): first, due to the presence of torsion the constraint contains terms with the second-order time derivative of spatial components, which can anyway be removed by means of field equations, and thus it is a real constraint that can then be plugged back into the field equations allowing them to specify the second-order time derivative of all components; second, the condition of complete antisymmetry of the spin admits only one independent contraction that eventually yields $V_{\rho}Q^{\rho\alpha\beta}=0$ and $W^{\nu}V^{\rho}=W^{\rho}V^{\nu}$, and therefore allowing us to write $Z_{\mu\nu}=\partial_{\mu}V_{\nu}-\partial_{\nu}V_{\mu}$, i.e. although we originally began with the differential operator given by the formal exterior derivative with respect to the most general connection we finally get the exterior derivative without additional fields: the consequence of this fact is that in this way the expression of the spin tensor can be inverted to let us write the torsion tensor as
\begin{eqnarray}
Q^{\rho\alpha\beta}=\frac{1}{2}
\left[V^{\alpha}\left(Z^{\rho\beta}+\frac{\lambda}{2}Z_{\sigma\theta}
\varepsilon^{\sigma\theta\rho\beta}\right)
-V^{\beta}\left(Z^{\rho\alpha}+\frac{\lambda}{2}Z_{\sigma\theta}
\varepsilon^{\sigma\theta\rho\alpha}\right)\right]
\end{eqnarray}
and equivalently the axial torsion vector as
\begin{eqnarray}
W_{\nu}=\frac{1}{6}\left(\lambda^{2}-1\right)V^{\alpha}Z^{\rho\beta}
\varepsilon_{\alpha\rho\beta\nu}
\end{eqnarray}
in terms of the vector field in the field equations allowing them to account for the back-reaction effects. Finally we have that $3\lambda W^{\nu}=(\lambda^{2}-1)Z^{\nu\rho}V_{\rho}$ is an important relationship between the axial torsion vector and the vector field.

First of all we easily see that in general cases in which the parameter $\lambda$ is different from zero, we have that we can substitute torsion through the torsion vector in terms of the vector field in the field equations, getting third-order derivatives within the field equations themselves; this problem can be solved by decomposing the vector field as $V_{\mu}=U_{\mu}+D_{\mu}B$ with $D_{\mu}U^{\mu}=0$ in terms of its transversal and longitudinal parts: the characteristic equation is given by 
\begin{eqnarray}
n^{2}\left(m^{2}+(\lambda^{2}-1)W^{2}\right)=(2\lambda^{2}-1)\left(n\cdot W\right)^{2}
\label{equation}
\end{eqnarray}
in terms of the torsion vector itself. 

Clearly this characteristic equations shows that to avoid time-like solutions to occur in the circumstance of weak torsion we have to require $2\lambda^{2}<1$ which expresses the fine-tuning of the parameter of the model; in this case we have that the condition of causality becomes $W^{2}<2m^{2}$ which is a condition expressing that torsion has a limit controlled from above by the mass of the vector field.

However, even restricting the discussion to the case in which the propagation is acceptable, we see that the condition of complete antisymmetry of the spin constitute a problem for the counting of degrees of freedom; indeed this condition accounts at least for an additional constraint that reduces the number of degrees of freedom to $2$ at most: this is not the right number of degrees of freedom possessed by the massive vector field.

We notice that for the particular case given by $\lambda^{2}\equiv0$ we do not get the characteristic equation (\ref{equation}) because in this case we have $Z^{\nu\rho}V_{\rho}\equiv0$ which gives $Z_{\mu\nu}Q^{\rho\mu\nu}=0$ and therefore the field equations reduce to those we would have had in absence of torsion; however, although torsion is not coupled to the vector field, it is nonetheless present with the condition of complete antisymmetry accounting for the additional constraint that reduces to $2$ the maximum number of degrees of freedom: even in this case the right amount of degrees of freedom is not achieved.

Finally, we consider the special case $\lambda^{2}\equiv1$ for which we have causality, which is due to the vanishing of torsion; however in this case too, although torsion would be zero and hence already completely antisymmetric, nevertheless the very vanishing of torsion is itself a condition that accounts for additional constraints which reduce to $2$ the maximum number of degrees of freedom: therefore in this case the right balance of degrees of freedom is not accomplished as well.

So, the most general Proca field can be fine-tuned to give a causal model, but all these causal models are overdetermined, and thus inconsistent.
\section*{Conclusion}
In this paper, we have considered the Proca vector field in which the dynamical term is written in terms of the curl or equivalently the exterior derivate, calculated with respect to the most general metric connection with completely antisymmetric Cartan torsion tensor; the most general system of field equations has been given, and discussed from the point of view of causal propagation and the geometrical properties constraining the degrees of freedom of the field.

It has been shown that the parameter $\lambda$ determines the features of the model: regarding the propagation, we proved that causality is ensured by the fine-tuning of the parameter given by $\lambda^{2}<\frac{1}{2}$, or else by $\lambda^{2}\equiv1$; regarding the geometric properties, we have seen that when $\lambda^{2}<\frac{1}{2}$ torsion undergoes the limitation given by $W^{2}<2m^{2}$, while for the special case $\lambda^{2}\equiv1$ the dynamical term is self-dual and torsion vanishes identically, so that in any case we have that the maximum value of torsion can never exceed a value given in terms of the mass of the field, and finally we have seen that this massive vector field is always overrestricted by the constraints arisen within the model. One point that should be stressed is that the special case $\lambda^{2}\equiv1$ is of fundamental interest because this special instance of self-dual dynamical term gives rise to a vanishing spin tensor, although a non-trivial spin tensor should be present in general for vector fields, and it is also intriguing that the occurrence of the condition $W^{2}<2m^{2}$ giving to torsion an upper value in terms of the mass of the field, which is a fact that admits no clear interpretation; on the other hand, the fact that this field never possesses the $3$ degrees of freedom that define massive vector fields constitutes an unsurmountable barrier. As this discussion has extensively underlined, any attempt to add the completely antisymmetric torsion to the metric connection in the exterior derivatives of the dynamical term for the Proca field implies inconsistencies.

So, albeit the inclusion of torsion could be a possible generalization for the Proca field, no such generalization actually leads to a consistent set of Proca field equations; therefore no such generalization gives rise to any consistent Proca field theory. 

This shows that it is already in its most general instance that the standard Proca theory is defined.

\end{document}